\documentclass[aps,twocolumn,superscriptaddress,notitlepage]{revtex4-1}
\usepackage{amssymb,amsfonts,amsmath,amstext,graphicx,hyperref}
\usepackage{tikz}
\usepackage{indentfirst}
\usepackage[capitalise]{cleveref}

\usetikzlibrary{quantikz}
\usetikzlibrary{angles,quotes}
\usepackage{hyperref}
\usepackage{graphicx}
\usepackage{parskip}
\usepackage[normalem]{ulem}
\usepackage{float}
\usepackage{blindtext}
\hypersetup{
	colorlinks=true,
	linkcolor=blue,
	filecolor=magenta,      
	urlcolor=cyan,
}
\usepackage{xfrac}
\usepackage{subfigure}
\definecolor{RoyalBlue}{cmyk}{1, 0.50, 0, 0}
\def\lket#1{\vert#1\rangle\hspace{-1mm}\rangle}

\def\ket#1{|{#1}\rangle}

\begin{document}

\title{Steady-State Tunable Entanglement Thermal Machine Using Quantum Dots}

\author{Anuranan Das}
\affiliation{Department of Electrical Engineering, Indian Institute of Technology-Bombay, Powai, Mumbai 400076, India}
\author{Adil Anwar Khan}
\affiliation{Department of Electrical Engineering, Indian Institute of Technology-Bombay, Powai, Mumbai 400076, India}
\author{Sattwik Deb Mishra}
\thanks{Current Address: E. L. Ginzton Laboratory, Stanford University, Stanford, CA 94305, USA}
\affiliation{Department of Electrical Engineering, Indian Institute of Technology-Bombay, Powai, Mumbai 400076, India}
\author{Parvinder Solanki}
\affiliation{Department of Physics, Indian Institute of Technology-Bombay, Powai, Mumbai 400076, India}
\author{Bitan De}
\affiliation{
Institute of Theoretical Physics, Jagiellonian University in Krak{\'o}w, Lojasiewicza 11, 30-348 Krak{\'o}w, Poland}
\author{Bhaskaran Muralidharan}
\affiliation{Department of Electrical Engineering, Indian Institute of Technology-Bombay, Powai, Mumbai 400076, India}
\author{Sai Vinjanampathy}
\email{sai@phy.iitb.ac.in}
\affiliation{Department of Physics, Indian Institute of Technology-Bombay, Powai, Mumbai 400076, India}
\affiliation{Centre for Quantum Technologies, National University of Singapore, 3 Science Drive 2, 117543 Singapore, Singapore}

\date{\today}

\begin{abstract}
We present a solid state thermal machine based on quantum dots to generate steady-state entanglement between distant spins. Unlike previous approaches our system is controlled by experimentally feasible steady state currents manipulated by dc voltages. By analyzing the Liouvillian eigenspectrum as a function of the control parameters, we show that our device operates over a large voltage region. As an extension, the proposed device also works as an entanglement thermal machine under a temperature gradient that can even give rise to entanglement at zero voltage bias. Finally, we highlight a post-selection scheme based on currently feasible non-demolition measurement techniques that can generate perfect Bell-pairs from the steady state output of our thermal machine.
\end{abstract} 
\maketitle

\section{Introduction}

\indent Quantum systems out of equilibrium play an important role in quantum transport \cite{dittrich1998quantum,dubi2011colloquium,wang2014nonequilibrium} and quantum thermodynamics \cite{millen2016perspective,goold2016role,vinjanampathy2016quantum}. Such quantum thermodynamic systems have focused on studying foundational issues such as large deviation statistics \cite{werner1992large,lebowitz2000large,vznidarivc2014exact,carollo2019unraveling}, the laws of thermodynamics \cite{kosloff2013quantum,vinjanampathy2016quantum,alicki2018introduction} and uncertainty relations \cite{miller2021thermodynamic}. Besides this, quantum thermodynamics has also provided useful implementations of quantum thermal machines including quantum engines \cite{Enginequan2005quantum,Enginequan2009quantum,Enginequan2009quantum,kosloff2014quantum,Enginecampisi2015nonequilibrium,noufal2020pre}, quantum refrigerators \cite{Ref_edwards1993quantum,Ref_rezek2009quantum,Ref_levy2012quantum,Ref_hofer2016autonomous,Ref_tan2017quantum}, diodes \cite{ordonez2017quantum} and amplifiers \cite{Amp_lambropoulos1967quantum,Amp_loudon1984properties,Amp_astafiev2010ultimate}. Numerous experiments have followed theoretical studies to elucidate various aspects of quantum thermal machines \cite{ExpRoss,ExpHuardDemon,ExpDeffner2018ErrorAnnealers,ExpMurch,ExpSerra,ExpPoem,ExpMainzFlywheel,Exp3ions,PhysRevA.100.042119,PhysRevLett.122.110601,PhysRevLett.125.166802,ExpCampisi2020DWave,ExpPekola,ExpMeasEngine,ExphenaoIBM,ExpCampisiFT,bouton2021quantum}.\\

\indent While typical quantum thermal machines output (mechanical or electrical) work, recently quantum thermal machines which output other quantities of interest have been discussed in the literature. For instance, entanglement thermal machines operate between two temperatures and output quantum states which are entangled\cite{Bohr_Brask_2015,tavakoli2018heralded,tavakoli2020autonomous}. Such entanglement thermal machines have been discussed with abstract models of finite and infinite dimensional systems \cite{aguilar2020entanglement} with potential implementations on a variety of physical platforms. To discuss entanglement generation we need to consider correlation between quantum systems. For instance, if a chain of quantum dots are coupled to each other and to an environment, the difference in the global and local master equations \cite{scali2021local} can highlight emergent correlations between different quantum dots. Likewise, if the system is outside the weak coupling regime \cite{strong_Coupling} steady states are expected to be coherent. Several quantum thermal machine proposals have used time dependent fields with varying experimental difficulty in implementation \cite{kohler2005driven,zhou2015boosting,landi2021non} to achieve entanglement between subsystems. For example, recent efforts have correlated quantum dots across millimetres \cite{pettas2020resonant}, though they require microwave mediation. What is more subtle to achieve is the entanglement of distant uncoupled impurity spins via their common interaction with a quantum dot. For more modest distances, an experimentally viable solid-state implementation of thermal machines that entangle distant spins and can be tuned by experimentally practical parameters such as gate voltages has been lacking thus far. In this manuscript, we remedy this by providing a gate or voltage tunable, steady state solid state entanglement switch. We also show broad stability regions of control parameters where the steady states are entangled. Unlike previous approaches towards controllable thermal machines, our proposal relies only on steady state currents (dc steady state) and voltages analyzed through Liouville eigenspectrum techniques \cite{suri2018speeding,gyamfi2020fundamentals,manzano2020short,solanki2021role} and quantum transport theory. \\

\begin{center}
\begin{figure}[h]
\subfigure[]{
\centering
\includegraphics[width=0.47\textwidth]{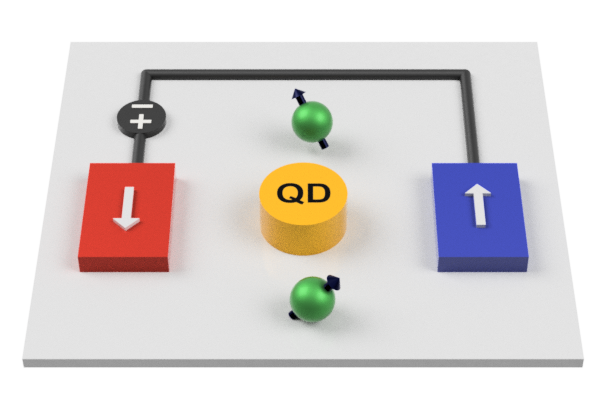}
\label{fig:Device3D}
}
\subfigure[]{
\centering
\includegraphics[width=0.41\textwidth]{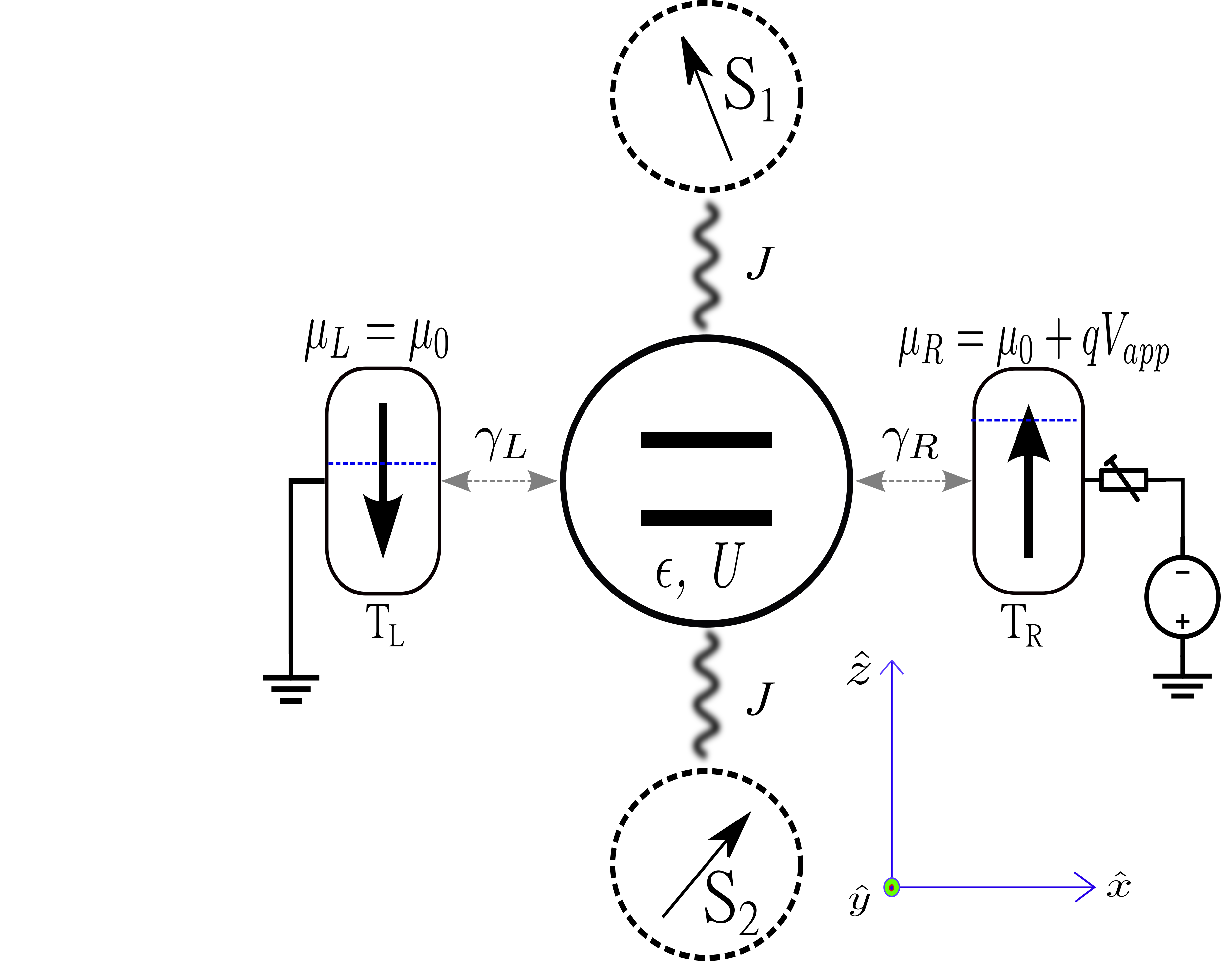}
\label{fig:Device2D}
}
\caption[]{Schematic of the voltage controlled steady state entanglement switch device : (a) The device consists of a quantum dot (shown in yellow) weakly coupled to spin polarized contacts (red and blue boxes) and placed in a bath of two spin \sfrac{1}{2} impurities (green spheres) with which it interacts strongly. Steady state entanglement is generated between the two constituent impurity spins which can be controlled by the voltage applied across the contacts. (b) The dot is coupled to oppositely spin polarized left and right contacts with electronic transport rates $\gamma_L$ and $\gamma_R$ respectively. Interaction coupling between the impurity spins and the quantum dot is represented by $J$. The left contact is held at constant potential $\mu_L$ and a voltage bias $V_{app}= (\mu_R-\mu_L)/q$ is applied across the two contacts which are at temperatures $T_L$ and $T_R$ respectively. Dot onsite energy is represented by $\epsilon$ and the charging energy is $U$. Interaction between the dot and bath is stronger as compared to the electronic transport rates and therefore the complete dot - impurity system can evolve and pass through entangled states before the electron exits.}
\label{fig:1}
\end{figure}
\end{center}

\indent Quantum transport theory provides a novel route toward realizing solid state platforms for quantum thermal machines and an alternate test bed for emerging ideas from quantum thermodynamics \cite{Datta,BM_Grifoni_1,BM_Grifoni_2,Sothmann_2014}. Specifically the phenomena of Coulomb blockade and spin blockade \cite{Datta,BM_blockade,BM_spinblock} provides a fertile platform for the realization of spin based qubits and their manipulation \cite{RevModPhys.79.1217,RevModPhys.85.961,10.1093/nsr/nwy153}. Additionally, quantum dot systems can act as quantum thermal machines via a voltage controlled thermoelectric setup \cite{Sothmann_2014,BM_Grifoni_1}. Our setup features a quantum dot connected to anti-parallel ferromagnetic contacts and a dc voltage applied across it can generate entanglement between two distant qubits connected independently to it. The setup presented in Fig.~\ref{fig:1} shows the quantum dot sandwiched between the two ferromagnetic contacts. The quantum dot is coupled to\\

\onecolumngrid

\vspace{-15mm}           
\twocolumngrid
\noindent two impurity spin qubits that are distant and not coupled to each other. Our analysis of the steady state shows that transport processes between the contacts and the quantum dot lead to entanglement between the two distant qubits, thus providing a steady state dc entanglement generating machine.   \\

\indent We begin with a brief description of the Hamiltonian of the quantum dot setup along with the impurity spins. After describing the transport setup, we present a brief review of how steady states are evaluated for Lindblad-type master equations. We then present various results indicating the tunability of the entanglement over a broad range of applied voltages as well as temperature gradients and discuss a post-selection scheme to generate perfect entanglement between two impurity spins.

\section{Model}

\indent The schematic of the entanglement switch setup is shown in Fig.~\ref{fig:Device3D}. The structure is a spin-valve based quantum dot, where the central block (in yellow) is a nonmagnetic quantum dot coupled to two ferromagnetic leads (represented by red and blue blocks) with opposite spin polarizations. In addition, the dot is coupled to two  magnetic impurities (in green) via exchange interaction and we assume a large Coulomb interaction at the impurity sites to hinder double occupancy in them. Functionally, this device generates entanglement between the spatially separated impurities (known as ancillae) by applying a dc bias across the magnetic leads when the transport through the dot is spin blocked \cite{BM_spinblock,Yoneda2020}. We begin by describing the Hamiltonians of different components associated with this setup. \\

\subsection{Hamiltonian of the setup}

\indent The composite Hamiltonian of the system is defined by $H=H_{D}+H_{C}+H_{A}+H_{DC}+H_{DA}$, where $H_{D}$, $H_{C}$ and $H_{A}$ represent the Hamiltonians of the dot, contacts and ancillae respectively. The term $H_{DC}$ is the tunneling Hamiltonian between the dot and contacts and $H_{DA}$ signifies the exchange coupling between the ancillae and the dot. The dot consists of two spin-degenerate energy levels at $\epsilon$ and with Coulomb interaction $U$ written as \cite{SET_book,anderson}
\begin{equation}
    H_{D}=\sum_\sigma \epsilon d_{\sigma}^{\dagger}d_{\sigma}+U d_{\uparrow}^{\dagger}d_{\uparrow}d_{\downarrow}^{\dagger}d_{\downarrow}.
    \label{dotH}
\end{equation}
Here $d_{\sigma}^{\dagger}$ ($d_{\sigma}$) is the creation (annihilation) operator of the electrons in the dot corresponding to the spin orientation in the dot basis $\sigma\in \uparrow, \downarrow$. The diagonalization of $H_D$ results in four energy levels at energies $0$, $\epsilon$, $\epsilon$ and $2\epsilon+U$ \cite{BM_blockade}. The magnetic leads $L,R$ represent  electronic reservoirs with non-interacting energy levels $\epsilon_{k,\sigma'}$ with crystal momentum $k$ and spin orientation $\sigma'$ defined in the contact basis. Therefore, $H_C$ reads as
\begin{equation}
    H_{C}=\sum_{\alpha(=L,R),k,\sigma'}\epsilon_{\alpha k \sigma'}c_{\alpha,k,\sigma'}^{\dagger}c_{\alpha,k,\sigma'}.
    \label{contactH}
\end{equation}
In the expression above, $c_{\alpha,k,\sigma'}^{\dagger}$ ($c_{\alpha,k,\sigma'}$) stands for the creation (annihilation) operators of the electrons in the contacts. The tunneling Hamiltonian $H_{DC}$ is defined as
\begin{equation}
        H_{DC}=\sum_{\alpha,k,\sigma'}\bigg[t_\alpha c_{\alpha,k,\sigma'}^{\dagger}d_{\sigma}+d_{\sigma}^{\dagger}c_{\alpha,k,\sigma'}\bigg]
        \label{tunnelH},
\end{equation}
where $t_\alpha$ represents the coupling energy between the electrons in the contacts and the dot and is independent of momentum $k$ and spin $\sigma'$. In numerous quantum dot spin valve structures, the spin direction of the magnetic leads are non-collinear \cite{kontos,SV_nature,SV_PRL}. If the polarization basis of the contact $\alpha$ with respect to the dot is tilted by angles $(\theta_\alpha, \phi_\alpha)$ (where $\theta_{\alpha}$ is the tilt with respect to the z-axis and $\phi_{\alpha}$ being the angle made with the azimuthal plane), then Eq.~\eqref{tunnelH} can be modified as \cite{koing2,koing1}
\begin{equation}
\begin{split}
 H_{DC}=\sum_{\alpha,k}\bigg[t_\alpha c_{\alpha,k,\uparrow}^{\dagger}\bigg(C_\alpha d_{\uparrow}+S_{\alpha}d_{\downarrow}\bigg)\bigg]+\\
        \sum_{\alpha,k}\bigg[t_\alpha c_{\alpha,k,\uparrow}^{\dagger}\bigg(-S_\alpha^{*} d_{\uparrow}+C_{\alpha}^{*}d_{\downarrow}\bigg)\bigg]+h.c, \end{split}
\label{tunnelH2}
\end{equation}
where the analytic expressions $C_\alpha=\cos({{\theta_\alpha}/{2}})exp({i{\phi_\alpha}/{2}})$ and $S_\alpha=\sin({{\theta_\alpha}/{2}})exp({-i{\phi_\alpha}/{2}})$ stand for the orthogonal components of dot spinor and h.c is the Hermitian conjugate. The dot-to-lead electron tunneling rate is defined as $\gamma_{\alpha \sigma'}=(2\pi/\hbar)|t_\alpha|^2 D_{\alpha \sigma'}$, where $D_{\alpha \sigma'}$ is the constant density-of-states of the magnetic leads $\alpha$ with spin orientation $\sigma'$. The polarization of lead $\alpha$ is defined by \cite{sothmann1,sothmann2} $p_\alpha=({\gamma_{\alpha,\uparrow}-\gamma_{\alpha,\downarrow}})/(\gamma_{\alpha,\uparrow}+\gamma_{\alpha,\downarrow})$. For the case in Fig.~\ref{fig:Device2D}, $p_L=-1$ and $p_R=+1$. The ancillae are magnetic impurities with energy $\epsilon_{a \sigma_a}$ and the Hamiltonian $H_A$ is defined in the limit with infinite Coulomb interaction as
\begin{equation}
    H_{A}=\sum_{a\in 1,2}\sum_{\sigma_a}\epsilon_{a \sigma_a}A_{\sigma_a}^{\dagger}A_{\sigma_a}
    \label{AH}.
\end{equation}
Here $\sigma_a\in \uparrow,\downarrow$ is the spin orientation in the ancilla basis and $A_{\sigma_a}^{\dagger}$ ($A_{\sigma_a}$) is the electronic creation (annihilation) operator in the ancilla. The exchange interaction between the dot and ancilla is given by
\begin{equation}
 H_{DA}=\frac{J}{\hbar^2}\bigg[S_D.(S_1+S_2)\bigg]
\label{DA}.   
\end{equation}
In Eq.~\eqref{DA}, $S_\nu,~\nu\in D,1,2$ are the total spin operators of the dot and ancillae and $J$ is the spin independent exchange coupling energy of the electrons in the dot and ancillae \cite{fazekas1999lecture}. In the spin basis of the dot, different components of spin operators are represented by the second-quantized notation as \cite{blundell2003magnetism} $S_D^{z}=\hbar(d_{\uparrow}^{\dagger}d_{\uparrow}-d_{\downarrow}^{\dagger}d_{\downarrow})/2$, $S_D^{x}=\hbar(d_{\uparrow}^{\dagger}d_{\downarrow}+d_{\downarrow}^{\dagger}d_{\uparrow})/2$ and $S_D^{y}=i\hbar(d_{\uparrow}^{\dagger}d_{\downarrow}-d_{\downarrow}^{\dagger}d_{\uparrow})/2$. In a similar manner, the spin operators of the ancilla $\alpha=1,2$ can be expressed by the fermionic operators as $S_{\alpha}^{z}=\hbar(A_{\alpha\uparrow}^{\dagger}A_{\alpha\uparrow}-A_{\alpha\downarrow}^{\dagger}A_{\alpha\downarrow})/2$, $S_{\alpha}^{x}=\hbar(A_{\alpha\uparrow}^{\dagger}A_{\alpha\downarrow}+A_{\alpha\downarrow}^{\dagger}A_{\alpha\uparrow})/2$ and $S_{\alpha}^{y}=i\hbar(A_{\alpha\uparrow}^{\dagger}A_{\alpha\downarrow}-A_{\alpha\downarrow}^{\dagger}A_{\alpha\uparrow})/2$. It is evident that the dot-to-ancilla interaction is finite only when both dot and ancillae are occupied by a single electron. Following the description of the Hamiltonians, our next goal is to formulate the time evolution equation of the reduced density matrix of the dot and ancillae. For the purpose of derivation, we use the transport formalism developed in \cite{Brower}, which yields the time evolution of the reduced density matrix $\rho$ in the form $\dot{\rho}=[R]\rho$ where $[R]$ stands for the transport rate matrix with the elements representing the contact assisted tunnelings between the states with differing electron numbers and the virtual precession of the degenerate states. In the subsequent stage, the matrix $[R]$ is utilized to compute the steady state using the  Liouville eigenspectra, which is then used to study entanglement generation.

\section{Transport in Quantum Dots}

\indent Our analysis closely follows the approach by Braig et.al., and other related works \cite{Brower,koing1, Grifoni_Non_Coll,Grifoni_SV,BM_Grifoni_2}. Firstly, we assume that the Coulomb-interactions in the ancilla impurities are high enough to forbid double occupancy of electrons inside them. Secondly, we assume that the strength of contact-to-dot electron coupling is smaller than the contact induced thermal broadening of the energy levels in the dot ($\gamma\ll kT/\hbar$). In this limit, the transport of electrons is carried out following the sequential-tunneling mechanism    \cite{beenakker_PRB,Timm,bitan_PRB,bitan_SCR,bitan_JPCM,bitan_JAP}, physically described by the quantum master equation. Due to the sequential tunnelings, the electron occupies the degenerate electronic states and leads to correlation between the states and the precession of spins \cite{BM_Grifoni_2}. Hence each of the time evolution equations contain two parts. Firstly the evolution equation contains a term for the sequential tunneling, which causes the increase or decrease of electron numbers in the system and secondly a virtual tunneling accounting for the precession of spins. \\

\onecolumngrid

\begin{figure}[H]
\centering
\subfigure[]{\label{fig:EnergySpectra}
\includegraphics[width=0.25\textwidth]{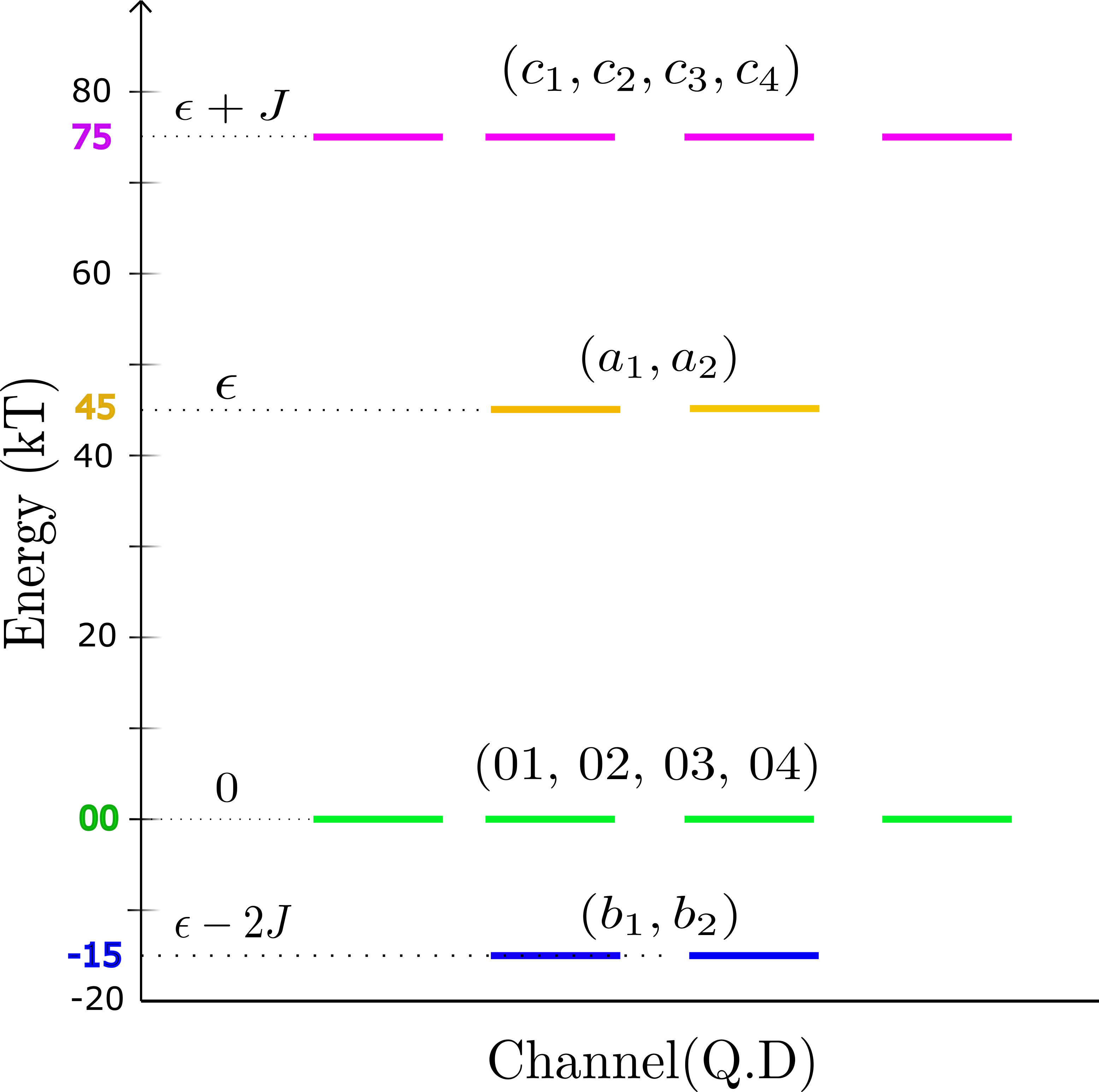}
}
\subfigure[]{\label{fig:TransitionSpectra}
\includegraphics[width=0.35\textwidth]{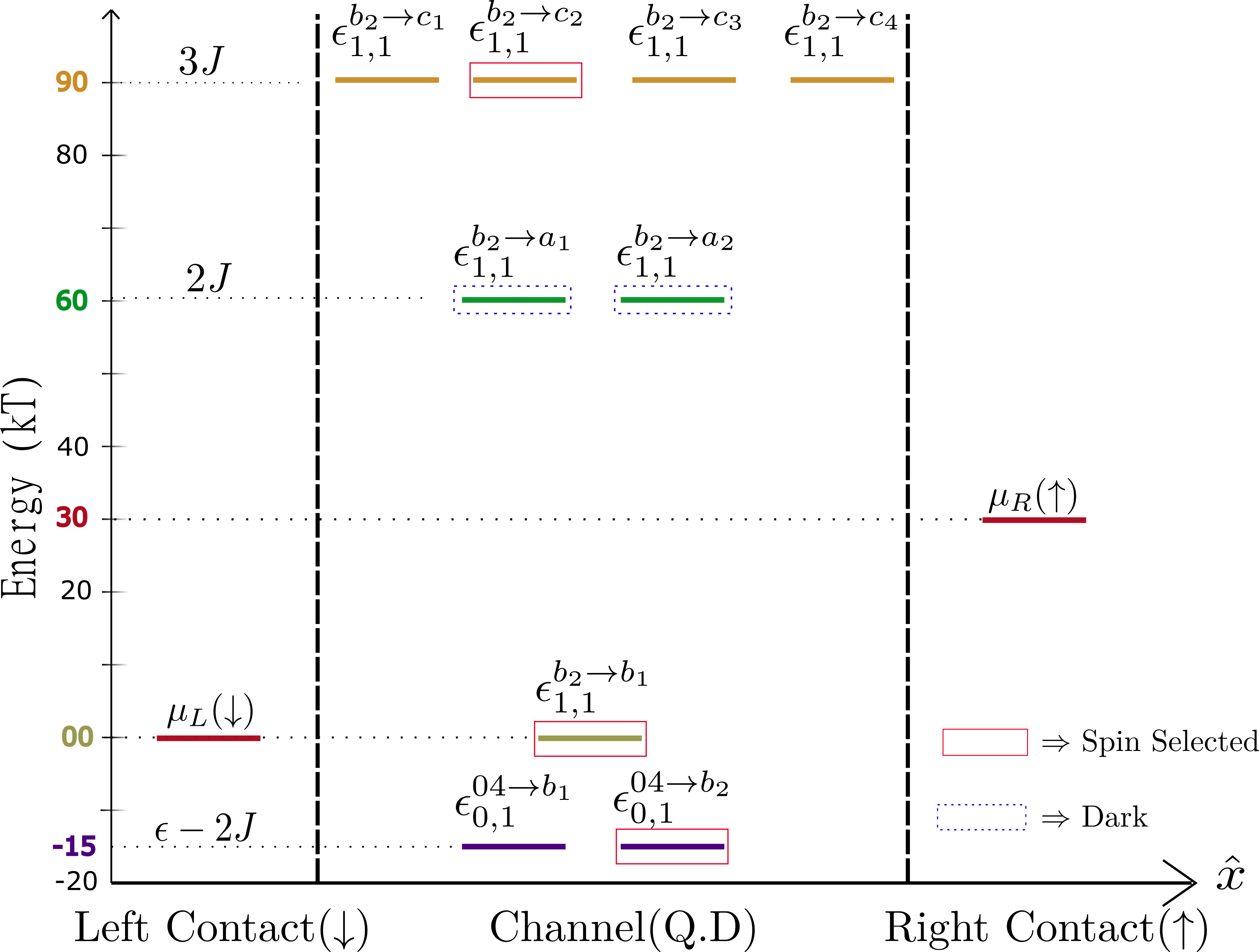}
}
\subfigure[]{\label{fig:Eigen}
\includegraphics[width=0.35\textwidth]{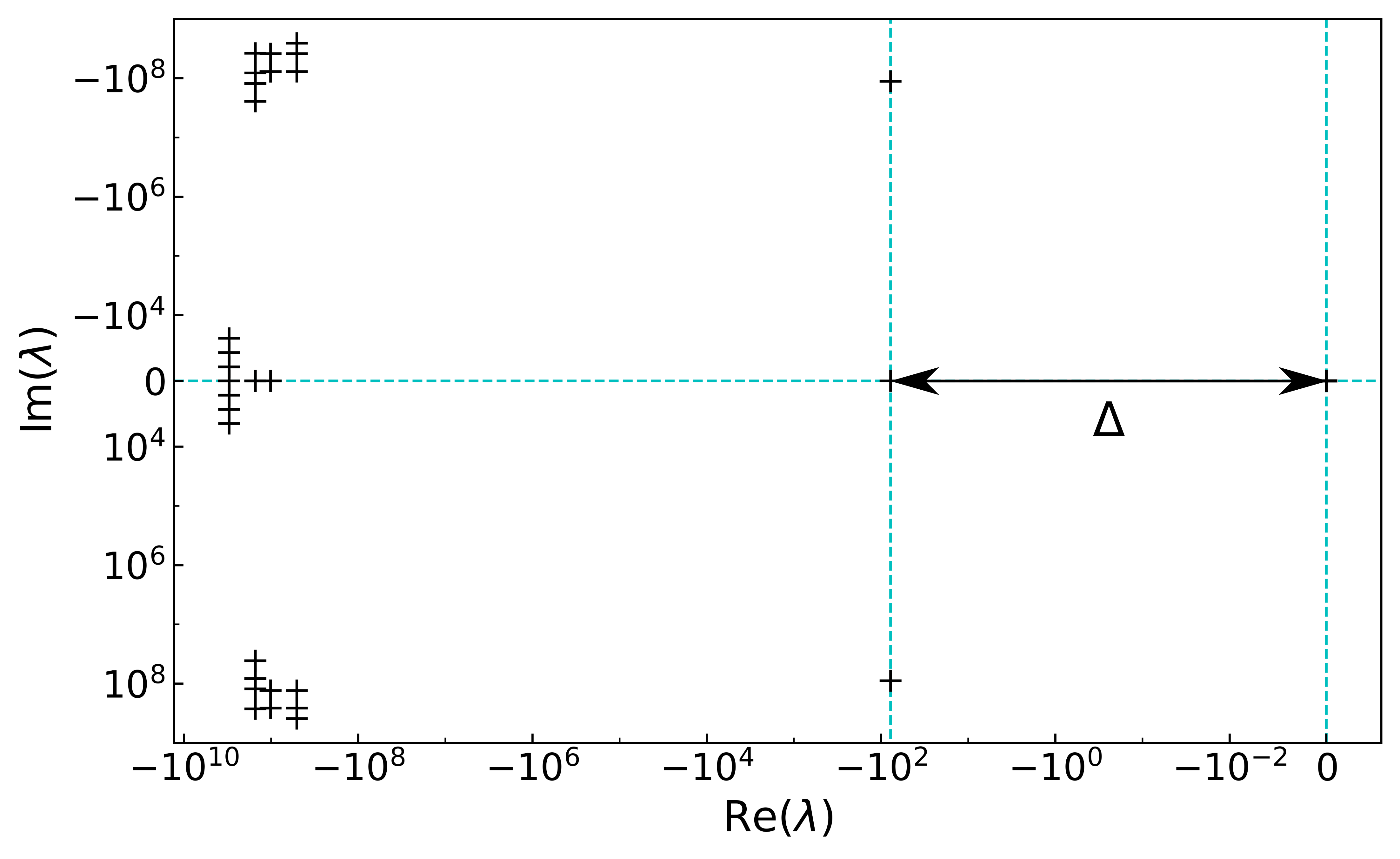}
}
\caption[]{Eigenspectra distribution for the Hamiltonian and Liouvillian dynamics at temperature of the contacts set to $T = 10$K with dot onsite energy $\epsilon = 45kT$, interaction coupling $J=30kT$ and the charging energy $U \to \infty$. The couplings to the reservoirs are
taken as $\hbar \gamma_{L}=\hbar \gamma_{R}=1 \mathrm{\mu eV}$. (a) Eigenenergies of the empty and one electron manifolds of the system Hamiltonian. (b) The viable transitions that are based on spin-selection rules are illustrated along with some dark transitions that do not occur (Not all dark transitions are shown). In this diagram, $\mu_L = \mu_0 = 0 kT$ and $\mu_R = 30 kT$. Hence the dot is kept biased at $V_{app} = 30 kT/e$. (c) Eigenvalue spectra of Liouvillian superoperator for given parameters values where eigenstate corresponding to zero eigenvalue is the steady state of the system.}
\label{fig:2}
\end{figure}
\twocolumngrid

\indent We  assume that the energy of the ancilla is $\epsilon_{a\sigma_a}=\mu=0$, where $\mu$ is the equilibrium chemical potential of the electronic contacts. As stated earlier, the ancillae are occupied by one electron each and hence the Hamiltonian of the system consisting of dot and ancilla can be diagonalized, producing sixteen eigenstates. Considering the polarization basis of the dot and ancilla to be parallel, four of these states $|01\rangle$, $|02\rangle$,$|03\rangle$ and $|04\rangle$ have energy 0 indicating a regime when the dot is empty, four states corresponding with double occupancy $|21\rangle$, $|22\rangle$,$|23\rangle$ and $|24\rangle$ have energy $2\epsilon+U$ and eight states with one electron in the dot. Among these eight states, there are two degenerate states $|a1\rangle, |a2\rangle$ with energy $\epsilon$, two states $|b1\rangle, |b2\rangle$ with energy $\epsilon-2J$ and four degenerate states $|c1\rangle,|c2\rangle,|c3\rangle,|c4\rangle$ with energy $\epsilon+J$. \\

\indent Now it is evident that the reduced density matrix of the dot and ancilla is a $16\times16$ block-diagonal matrix with $4\times4$ block  $\rho_0$ for states with empty dot, $8\times8$ block $\rho_1$ for states with one electron occupied dot and $4\times4$ block $\rho_2$ for states with double occupied dot. Further, $\rho_1$ is another block diagonal matrix, first with $\rho_{1,a}$ as $2\times2$ block for degenerated states $|ai\rangle$ ($i\in 1,2$), then $\rho_{1,b}$ as $2\times2$ block for degenerated states $|bj\rangle$ ($j\in 1,2$) and lastly the $\rho_{1,c}$ $4\times4$ block for degenerated states $|ck\rangle$ ($k\in 1,2,3,4$). Following trace preservation, the block matrices must obey  $tr \rho_0+tr \rho_{1,a}+tr \rho_{1,b}+tr \rho_{1,c}+tr \rho_2=1$, the diagonal terms of $\rho$ signify the probability of the respective eigenstates and the off-diagonal terms represent the intra-state coherences.\\
\begin{table}[!h]
\caption{The sixteen eigenstates obtained on diagonalizing the system Hamiltonian} 
\begin{ruledtabular}
\begin{tabular}{ccl}
Notation&State&Energy\\
\colrule
$\ket{01}$ & $\ket{0} \otimes \ket{\uparrow\uparrow}$ & $0$\\
$\ket{02}$ & $\ket{0} \otimes \ket{\uparrow\downarrow}$ & $0$\\
$\ket{03}$ & $\ket{0} \otimes \ket{\downarrow\uparrow}$ & $0$\\
$\ket{04}$ & $\ket{0} \otimes \ket{\downarrow\downarrow}$ & $0$\\
$\ket{a1}$ & $\ket{\downarrow} \otimes \ket{\Psi^{-}}$ & $\epsilon$\\
$\ket{a2}$ & $\ket{\uparrow} \otimes \ket{\Psi^{-}}$ & $\epsilon$\\
$\ket{b1}$ & $\sqrt{\frac{1}{3}}\ket{\uparrow} \otimes \ket{\Psi^{+}} - \sqrt{\frac{2}{3}}\ket{\downarrow} \otimes \ket{\uparrow \uparrow}$ & $\epsilon-2J$\\
$\ket{b2}$ & $\sqrt{\frac{1}{3}}\ket{\downarrow} \otimes \ket{\Psi^{+}} - \sqrt{\frac{2}{3}}\ket{\uparrow} \otimes \ket{\downarrow \downarrow}$ & $\epsilon-2J$\\
$\ket{c1}$ & $\ket{\downarrow} \otimes \ket{\downarrow\downarrow}$ & $\epsilon+J$\\
$\ket{c2}$ & $\ket{\uparrow} \otimes \ket{\uparrow\uparrow}$ & $\epsilon+J$\\
$\ket{c3}$ & $\sqrt{\frac{2}{3}} \ket{\uparrow} \otimes \ket{\Psi^{+}} + \sqrt{\frac{1}{3}} \ket{\downarrow} \otimes \ket{\uparrow \uparrow}$ & $\epsilon+J$\\
$\ket{c4}$ & $\sqrt{\frac{2}{3}} \ket{\downarrow} \otimes \ket{\Psi^{+}} + \sqrt{\frac{1}{3}} \ket{\uparrow} \otimes \ket{\downarrow \downarrow}$ & $\epsilon+J$\\
$\ket{21}$ & $\ket{\uparrow\downarrow}\otimes\ket{\uparrow\uparrow}$ & $2\epsilon+U$\\
$\ket{22}$ & $\ket{\uparrow\downarrow}\otimes\ket{\uparrow\downarrow}$ & $2\epsilon+U$\\
$\ket{23}$ & $\ket{\uparrow\downarrow}\otimes\ket{\downarrow\uparrow}$ & $2\epsilon+U$\\
$\ket{24}$ & $\ket{\uparrow\downarrow}\otimes\ket{\downarrow\downarrow}$ & $2\epsilon+U$\\
\end{tabular}
\end{ruledtabular}
\label{tab:ConcurrenceEvolution}
\end{table}

\indent The tunneling Hamiltonian for each contact $\alpha$ is a $16\times 16$ matrix with matrix elements $w_{\alpha,q,q'}^{\pm}$, where $|q\rangle$ and $|q'\rangle$ are the electronic states associated with tunneling. In the case of sequential tunneling, $w_{\alpha,q,q'}^{\pm}=0$ unless the electron numbers in the states $N_{q}=N_{q'}\pm 1$. The magnitude of $w_{\alpha,q,q'}^{\pm}$ is closely dependent on the electronic wave function of the states $|q\rangle$ and $|q'\rangle$ \cite{wf}. In the presence of degeneracy, the magnitude of $w_{\alpha,q,q'}^{\pm}$ is a function of Clebsch-Gordon coefficients associated with different degenerate states. For the device shown in Fig.~\ref{fig:1}, the energies of different electronic states in the joint basis of the dot and ancillae are depicted in Fig.~\ref{fig:EnergySpectra}. In addition, the corresponding expressions of their eigenstates are tabulated in Table~\ref{tab:ConcurrenceEvolution}. Apart from the condition $N_{q}=N_{q'}\pm 1$, the sequential tunneling of electrons between these states $|q\rangle$ and $|q'\rangle$ is only feasible if the energy difference between them $\epsilon^{\ket{q}}_{N_{q}}-\epsilon^{\ket{q'}}_{N_{q'}}$ falls within the transport window opened up by the chemical potential of the leads $\mu_R-\mu_L$. For the sake of illustration, we present a few transitions based on energy difference and spin selection via Fig.~\ref{fig:TransitionSpectra}, where the following notation is used to describe the allowed sequential transitions in the system:
\begin{equation}
    \epsilon^{\ket{q} \rightarrow \ket{q'}}_{N_q, N_{q'}} = \epsilon^{\ket{q}}_{N_q} - \epsilon^{\ket{q'}}_{N_q'}.
    \label{eq:transition_difference}
\end{equation}
\indent Besides sequential tunneling, the magnetic polarizations of the contacts $\alpha$ and the degeneracy of the states $|q\rangle$ and $|q'\rangle$ cause virtual precession within the degenerate space of the dot Hamiltonian. 
For state $|q\rangle$,  this can be described by an effective Hamiltonian \cite{gurvitz1998rate,virtual1,von2001spectroscopy,aleiner2002quantum}
\begin{equation}
	\begin{split}
		H_{q}=\frac{\hbar}{4\pi}\mathcal{P}\int dE \sum_{\alpha,q'}^{}\bigg(1-2f_{\alpha}(E)\bigg)\\
		\times\bigg[\frac{\Gamma_{\alpha,q,q'}^{+}\Gamma_{\alpha,q,q'}^{-}}{\epsilon_{q}-\epsilon_{q'}-E}\delta(N_q-N_{q'}-1)+\\\frac{\Gamma_{\alpha,q,q'}^{-}\Gamma_{\alpha,q,q'}^{+}}{\epsilon_{q'}-\epsilon_{q}-E}\delta(N_q-N_{q'}+1)\bigg],
		\label{eq:virtual}
	\end{split}
\end{equation}
where $\mathcal{P}$ denotes the Cauchy principal value, $\delta$ stands for the Dirac delta function and $\Gamma_{\alpha,q,q'}^{\pm}=\sqrt{\gamma_{\alpha}}w_{\alpha,q,q'}^{\pm}$ \cite{Brower}. The virtual tunneling processes are a function of the Fermi-Dirac distribution function $f_{\alpha}$ associated with contact $\alpha$, namely
\begin{equation}
\centering
	f_{\alpha}(E)=\frac{1}{1+exp(\frac{E-\mu_{\alpha}}{kT_{\alpha}})},
\end{equation}
where $\mu_{\alpha}$ and $T_{\alpha}$ are the chemical potential and temperature of the electronic contact $\alpha$. Now one can derive the time evolution equations for different blocks of density matrix of the system as follows \cite{Brower,albert_PRB} to obtain
\begin{equation}
	\begin{gathered}
	\frac{d\rho_{0n}}{dt}=\frac{i}{\hbar}[\rho_{0n},H_{0n}]\\
	+\sum_{\alpha,{1n'}}\bigg[(1-f_{\alpha}(\epsilon_{1n'}-\epsilon_{0n}))\Gamma_{\alpha,0n,1n'}^{-}\rho_{1n'}\Gamma_{\alpha,0n,1n'}^{+}\\
	-f_{\alpha}(\epsilon_{1n'}-\epsilon_{0n})\Gamma_{\alpha,0n,1n'}^{-}\rho_{0n}\Gamma_{\alpha,0n,1n'}^{+}\bigg].
	\end{gathered}
	\label{eq:drho_0/dt}
\end{equation}
\begin{equation}
	\begin{gathered}
		\frac{d\rho_{1n'}}{dt}=\frac{i}{\hbar}[\rho_{1n'},H_{1n'}]\\
		+\sum_{\alpha,{0n}}\bigg[-\frac{1}{2}\big(1-f_{\alpha}(\epsilon_{1n'}-\epsilon_{0n})\big)[\Gamma_{\alpha,0n,1n'}^{+}\Gamma_{\alpha,0n,1n'}^{-}\rho_{1n'}\\
		+\rho_{1n'}\Gamma_{\alpha,0n,1n'}^{+}\Gamma_{\alpha,0n,1n'}^{-}]\\
		+(1-f_{\alpha}(\epsilon_{1n'}-\epsilon_{0n}))\Gamma_{\alpha,0n,1n'}^{+}\rho_{0n}\Gamma_{\alpha,0n,1n'}^{-}\bigg]\\
		-\sum_{\alpha,{2n''}}\bigg[\frac{1}{2}f_{\alpha}(\epsilon_{2n''}-\epsilon_{1n'})[\Gamma_{\alpha,1n',2n''}^{-}\Gamma_{\alpha,0n,1n'}^{+}\rho_{1n'}\\
		+\rho_{1n'}\Gamma_{\alpha,0n,1n'}^{-}\Gamma_{\alpha,0n,1n'}^{+}]\\
		+\big(1-f_{\alpha}(\epsilon_{2n''}-\epsilon_{1n'})\big)\Gamma_{\alpha,1n',2n''}^{-}\rho_{2n''}\Gamma_{\alpha,1n',2n''}^{+}\bigg].
	\end{gathered}
	\label{eq:drho_1/dt}
\end{equation}
\begin{equation}
	\begin{gathered}
		\frac{d\rho_{2n''}}{dt}=\frac{i}{\hbar}[\rho_{2n''},H_{2n''}]\\+\sum_{\alpha,{1n'}}\bigg[f_{\alpha}(\epsilon_{2n''}-\epsilon_{1n'}))\Gamma_{\alpha,1n',2n''}^{+}\rho_{1n'}\Gamma_{\alpha,1n',2n''}^{-}\\-\big(1-f_{\alpha}(\epsilon_{2n''}-\epsilon_{1n'})\big)\Gamma_{\alpha,1n',2n''}^{+}\rho_{2n''}\Gamma_{\alpha,1n',2n''}^{-}\bigg].
	\end{gathered}	
	\label{eq:drho_2/dt}
\end{equation}
\indent In the equations above $0n\in 01,02,03,04$, $2n''\in 21,22,23,24$ and $1n'\in a1,a2,b1,b2,c1,c2,c3,c4$. By considering the equations \cref{eq:drho_0/dt,eq:drho_1/dt,eq:drho_2/dt}, the time evolution equation of the reduced density matrix of the dot and ancillae is given by $\dot{\rho}=[R]\rho$ with $[R]$ being the transport rate matrix and $[R]$ is used to compute the Liouvillian superoperator in the subsequent stages. From the transport perspective, it is imparative to check the conservation of charge and hence the electron current enamating from the leads $L$ and $R$ should counterbalance $I_L=-I_R$.  This indeed is the case. The expression of current associated with contact $\alpha\in L,R$ is derived as
\begin{equation}
	\begin{gathered}
		I_{\alpha}=-|e|\sum_{0n,1n',2n''}^{}f_{\alpha}(\epsilon_{1n'}-\epsilon_{0n})\Gamma_{\alpha,0n,1n'}^{-}\rho_{0n}\Gamma_{\alpha,0n,1n'}^{+}-\\
		\bigg(1-f_{\alpha}(\epsilon_{1n'}-\epsilon_{0n})\bigg)\Gamma_{\alpha,0n,1n'}^{-}\rho_{1n'}\Gamma_{\alpha,0n,1n'}^{+}+\\
		f_{\alpha}(\epsilon_{2n''}-\epsilon_{1n'}))\Gamma_{\alpha,1n',2n''}^{+}\rho_{1n'}\Gamma_{\alpha,1n',2n''}^{-}
		\\-\bigg(1-f_{\alpha}(\epsilon_{2n''}-\epsilon_{1n'})\bigg)\Gamma_{\alpha,1n',2n''}^{+}\rho_{2n''}\Gamma_{\alpha,1n',2n''}^{-}.
	\end{gathered}
\end{equation}

\indent It is evident that for our system the dimension of the matrix $[R]$ is $16\times16$ and we convert it to Liouville superoperator $\mathcal{L}$ of dimension $256\times 256$. In the next section, we utilize $\mathcal{L}$ to compute the steady state density matrix.

\section{Steady States of Lindblad-type Master Equation}

\indent The master equation described by \cref{eq:drho_0/dt,eq:drho_1/dt,eq:drho_2/dt}
are of Lindblad form, written canonically as
\begin{equation}
    \Dot{\rho}=-i[H,\rho]+\sum_k \gamma_k\mathcal{D}[O_k]\rho.
    \label{eq:master}
\end{equation}
Here $H$ represents a generic Hamiltonian component of the evolution and $\mathcal{D}[O_k]\rho:=O_k\rho O_k^{\dagger}-\frac{1}{2}\{ O_k^{\dagger} O_k,\rho\}$ represent the Lindblad operators acting on the density matrix $\rho$. The transient and steady state dynamics of such an open system evolution can be inferred from the eigenspectra of the Liouville superoperator. Such a Liouville superoperator vectorizes the dynamics represented in Eq.~\eqref{eq:master} to Liouville-Schmidt space, given by the form
\begin{equation}
    \lket{\dot{\rho}}=\mathcal{L}\lket{\rho}.
    \label{eq:L_master}
\end{equation} 
Here $\mathcal{L}$ is the Liouvillian superoperator \cite{manzano2020short,suri2018speeding,li2014perturbative,solanki2021role} and $\lket{\rho}$ is the vector representation of matrix $\rho$. Eigenvector $\lket{\rho}$ is obtained by stacking the columns of density matrix $\rho$ while $\mathcal{L}$ is obtained by the related transformation $A\rho B\rightarrow B^* \otimes A \lket{\rho}$ \cite{manzano2020short,suri2018speeding}. Following the given transformation, mathematical form of $\mathcal{L}$ is given by 
\begin{eqnarray}
    \mathcal{L}&=&-i(I\otimes H-H^*\otimes I)+\sum_k\gamma_k(O_k^T\otimes O_k \\ \nonumber &-&\frac{1}{2}[I\otimes O_k^{\dagger}O_k+O_k^T O_k^* \otimes I]).
\end{eqnarray}
The Liouville superoperator $\mathcal{L}$ is generally non-Hermitian and has complex eigenvalues ($\lambda_k=\alpha_k+i\beta_k$) with different left($u_k$) and right eigenvectors($w_k$) defined by $ \mathcal{L}\lket{u_k}=\lambda_k\lket{u_k}$ and $\mathcal{L}^{\dagger}\lket{w_k}=\lambda_k^*\lket{w_k}.$
The solution of Eq.~\eqref{eq:L_master} at any given time $t$ can be written as
$\lket{\rho(t)}=e^{\mathcal{L}t}\lket{\rho_0}$ where $\lket{\rho_0}$ is the initial state. 
Physical density matrices are positive, trace class and Hermitian \cite{nielsen2002quantum}. These properties are preserved by the Liouville superoperator by the condition $\alpha_k\leq 0~\forall~k$. Furthermore, $\beta_k$ always appear as complex conjugate pairs. Both  of these properties can be observed in the eigenspectrum of the Liouville operator corresponding to \cref{eq:drho_0/dt,eq:drho_1/dt,eq:drho_2/dt} shown in Fig.~\ref{fig:Eigen}. Steady state solution of Eq.~\eqref{eq:L_master} can be written as $\lket{\rho_{ss}}=\lim_{t\rightarrow \infty}e^{\mathcal{L}t}\lket{\rho_0}$. Since eigenvectors corresponding to eigenvalues having negative real part will dissipate out at $t\rightarrow \infty$, steady states correspond to eigenvectors of $\mathcal{L}$ with zero eigenvalue. In the presence of symmetries, $\mathcal{L}$ can have multiple steady states. The numerical diagonalization of the Liouville superoperator in this case can render unphysical solutions, which are a consequence of rotations in the symmetry subspace. This can be accounted for by imposing positivity in the diagonalisation procedure \cite{thingna2021degenerated}.\\

\section{Results}

\indent The coupling rate between the dot and contact ($\gamma_{\alpha}$) and the lead polarization directions are considered to be fixed in this model. On the other hand, tunable parameters include (i) the voltage bias (applied between the contacts) ($V_{app}$), (ii) the temperatures of the contacts ($T_L$ and $T_R$,) (iii) the dot onsite energy ($\epsilon$) and (iv) the exchange coupling $(J)$. The value of the dot onsite energy can be tuned using a gate bias. In the case of tunnel coupling between the quantum dot and the impurity states, one can even use a separate gate bias to the change the magnitude of the exchange coupling between the \cite{RevModPhys.79.1217,RevModPhys.85.961} dot and the impurity spins. At any given value of the tunable parameters $V_{app}$, $T_L$, $T_R$, $\epsilon$ and $J$, we can diagonalize the Liouville superoperator corresponding to the master equation in Eq.~\eqref{eq:master} to understand the steady state entanglement properties of the system. Based on an appropriate choice of these parameters, we can make the unperturbed system evolve with time into different steady states. The dependence of the steady state and hence the concurrence of the system on these four parameters is illustrated in Fig.~\ref{fig:3} which we discuss in detail with respect to variation of individual controls.

\subsection{Variation of the applied voltage}

\begin{figure}[H]
\center
\subfigure{\label{fig:pvsMu}
\includegraphics[width=0.45\textwidth]{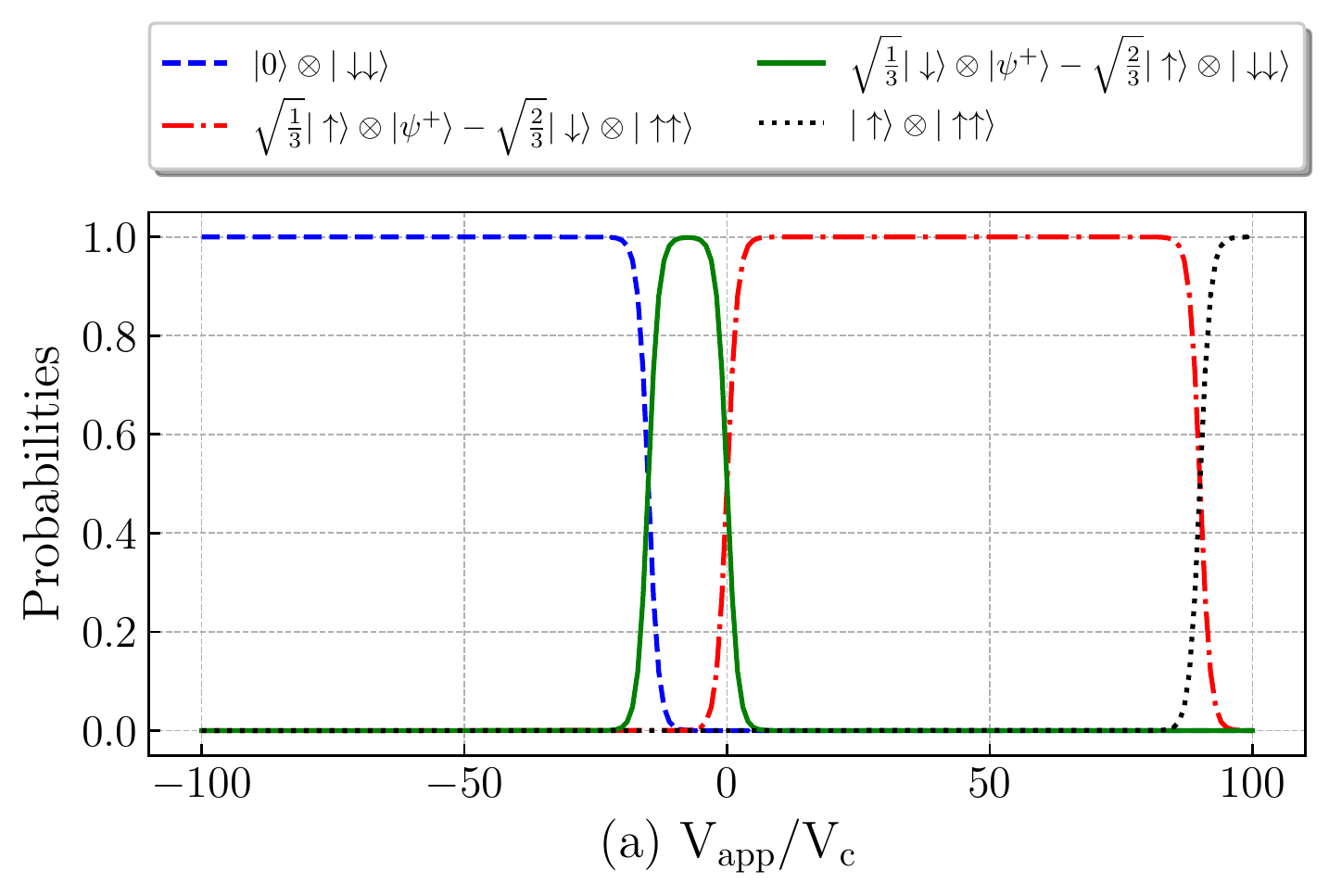}
}
\subfigure{\label{fig:CvsMu}
\includegraphics[width=0.45\textwidth]{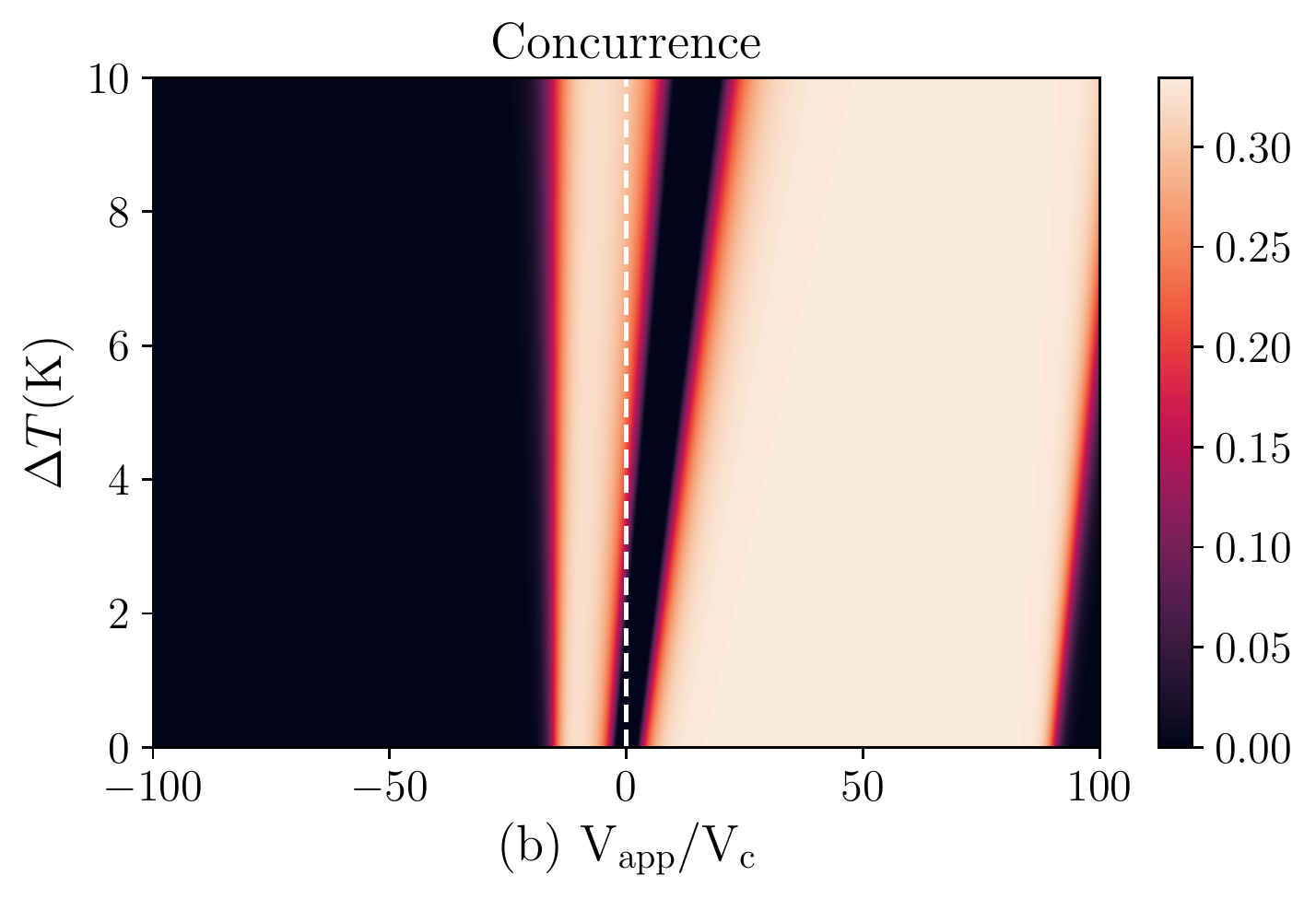}
}
\subfigure{\label{fig:Heatmap_w}
\includegraphics[width=0.45\textwidth]{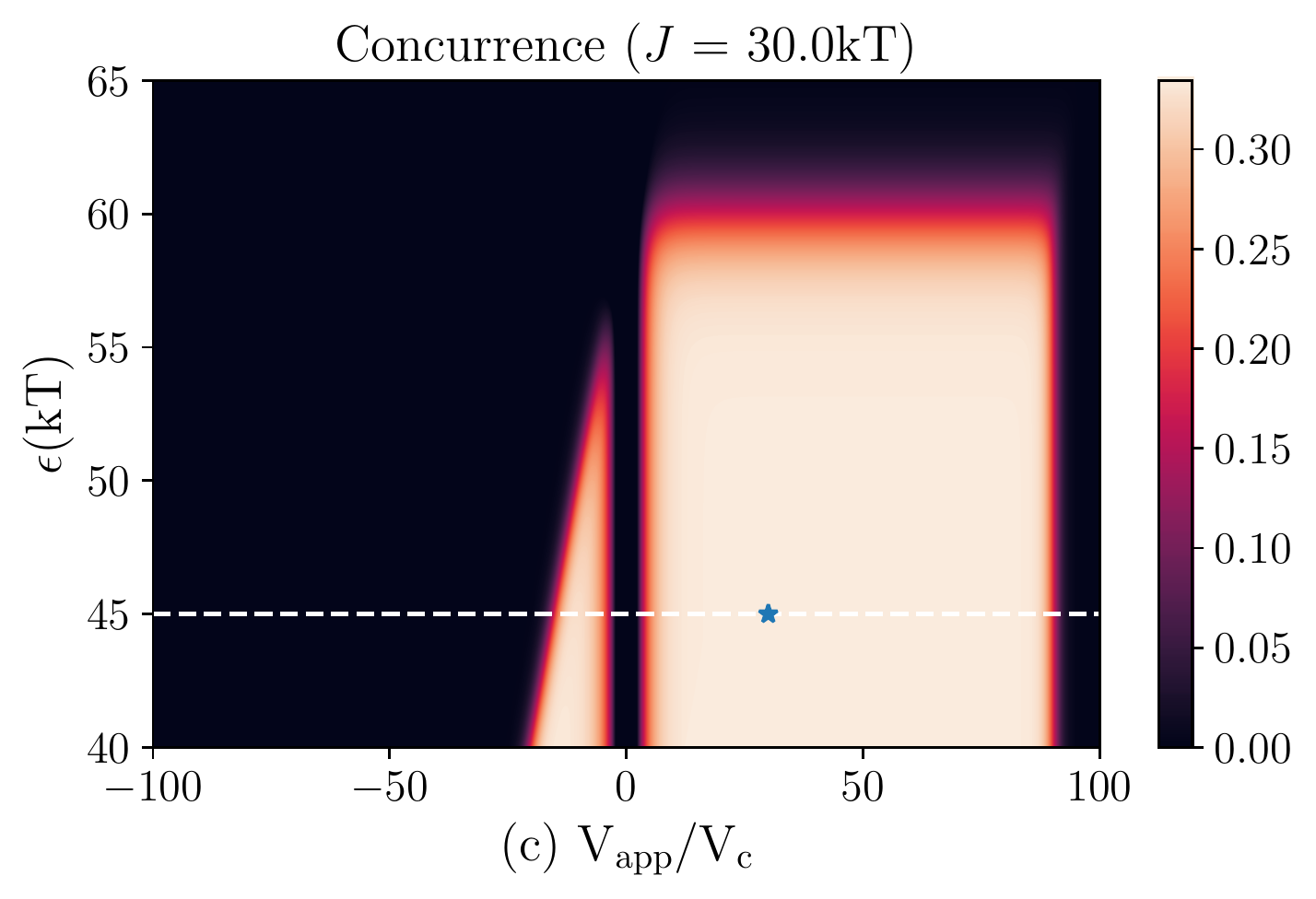}
}
\subfigure{\label{fig:Heatmap_J}
\includegraphics[width=0.45\textwidth]{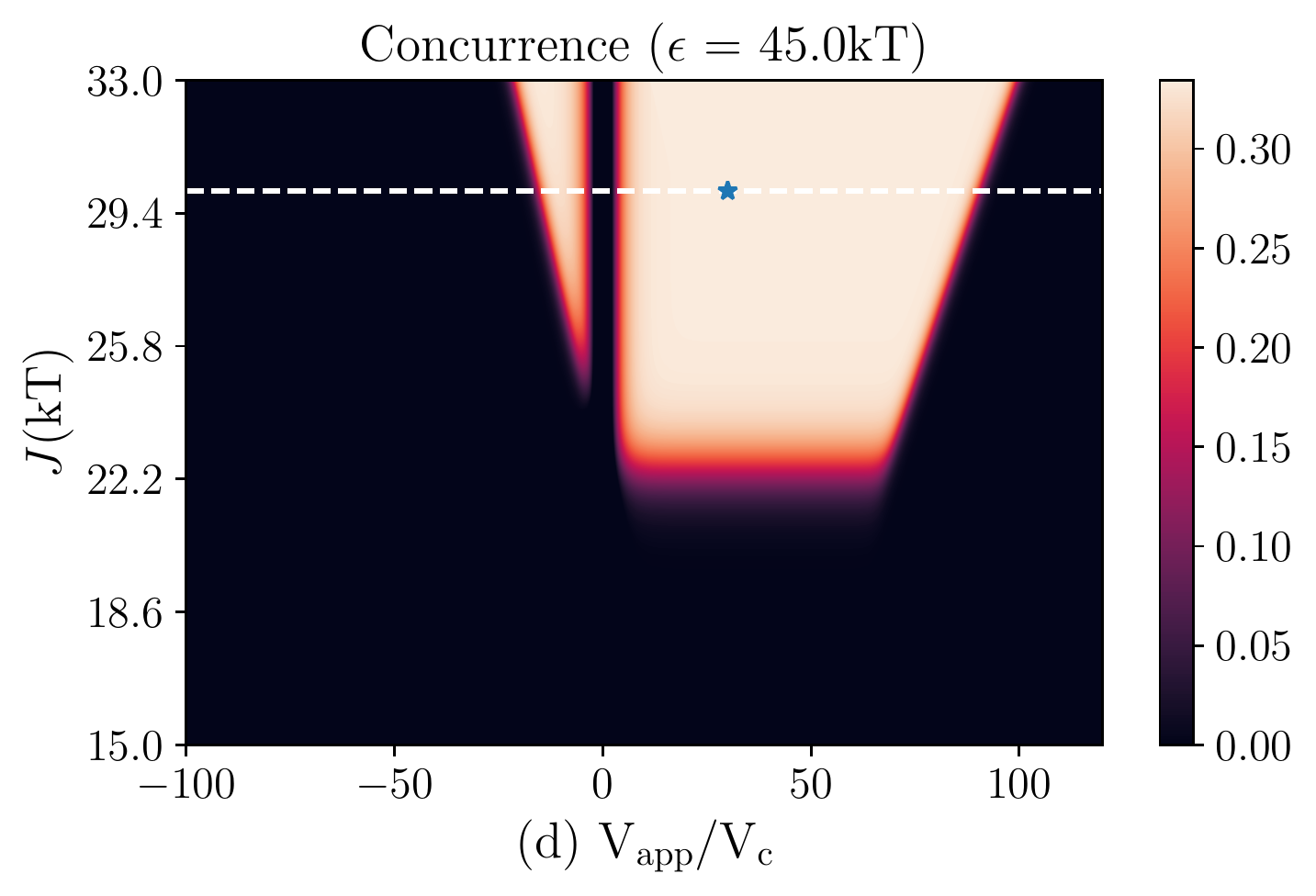}
}
\caption[]{Concurrence of the ancilla qubits under steady state for various parameters ($V_c = kT/e$). (a) Probability of steady state occupation of eigenstates with voltage sweep applied to the contacts at $\epsilon = 45 kT$, $J = 30 kT$ and $T = T_L = T_R$ = 10K. (b) Variation of concurrence with temperature gradient $\Delta T = T_R - T_L$ and applied voltage (while $\epsilon = 45 kT_L$ and $J = 30 kT_L$). The dashed white line is drawn along zero voltage bias. (c) Variation of concurrence with the interaction coupling $J$ varied while $\epsilon = 45 kT$. (d) Variation of concurrence with the dot onsite energy $\epsilon$ varied with $J= 30 kT$. In both (c) and (d) $T = T_L = T_R$ = 10K. The dashed white lines in (c) and (d) represent the cut section corresponding to (a). The asterisk (*) corresponds to the conditions under which Liouvillian discussed in Fig.~\ref{fig:2} is set.}
\label{fig:3}
\end{figure}

\indent Figure~\ref{fig:pvsMu} shows the steady state probabilities of different eigenstates with applied voltage. To demonstrate entanglement generation, we consider the other two parameters to be the same as in Fig.~\ref{fig:2}, namely $\epsilon = 45 kT$, $J = 30 kT$ and $T = T_L = T_R$ = 10K. The values of these parameters were chosen in correspondence with theoretically calculated values for modern nanoscale devices \cite{RevModPhys.79.1217,RevModPhys.85.961}. In our model, we consider the left contact to be fixed at zero and the electrochemical potential of the right contact moves with the applied voltage bias. Additionally, since the exchange interaction energy ($J$) is positive, therefore an antiferromagnetic spin alignment, that is based on the the $S_z$ and the total $S$ symmetry \cite{BM_spinblock,Ghosh_BM} between the dot and the ancilla electrons becomes energetically favorable. \\

\indent Four distinct regions are observed as the voltage is varied in Fig.~\ref{fig:pvsMu}. Two of these regions, characterized by $V<-15V_c$ and $V>90V_c$ represents the zero occupancy states of the dot with fully polarized impurity spins where $V_c=kT/e$. The other two regions, characterized by $-15V_c<V<0$ and $15V_c\leq V<90V_c$ represent an unequal superposition of single occupancy spin states in the quantum dot accompanied by entangled or product states in the ancilla. For example, the quantum state in the region $15V_c\leq V<90V_c$ is $\sqrt{1/3}\ket{\uparrow}\otimes\ket{\psi^+}-\sqrt{2/3}\ket{\downarrow}\otimes\ket{\uparrow\uparrow}$.
It can be observed that there are three steady state transitions that take place, one each at $-15kT$, $0kT$ and $90kT$, corresponding to $\epsilon - 2J$, $2J$ and $3J$ respectively. These spin selected transitions were shown in Fig.~\ref{fig:TransitionSpectra}.  
\onecolumngrid

\twocolumngrid
\noindent We also note a rise in the expectation value of the z-component $\langle S_z \rangle$ of total spin as the right contact voltage is increased, for example on transition from $\ket{b2}$ to $\ket{b1}$ around zero bias, $\langle S_z \rangle$ changes from -1/2 to +1/2. This aspect is a result of voltage driven ``precession'' \cite{Brower,koing1,Grifoni_Non_Coll,Grifoni_SV,BM_Grifoni_2} in the degenerate subspace of the states $b1$ and $b2$, which typically happens via the virtual transitions represented in Eq.~\eqref{eq:virtual}.

\subsection{Variation of Temperature Gradient ($\Delta T$)}

\indent Alongside the application of a voltage, described as a voltage controlled thermoelectric \cite{BM_Grifoni_1,Sothmann_2014,bitan_JAP}, the setup can also function as a thermal machine when subject to a temperature gradient. The application of a temperature gradient typically gives rise to a closed circuit current and an open circuit voltage (Seebeck voltage), resulting in a region for useful energy conversion. In typical thermoelectric applications, one seeks to maximize the output power at a given efficiency, which is related to the output currents at zero voltage and the open circuit voltages. As a result of this thermoelectric effect, interestingly, we can see that our device gives rise to entanglement at zero bias voltage. We can clearly notice that the entanglement profile shifts toward zero bias as the temperature gradient is increased, thereby leading to a sizable entanglement generation at zero bias  as indicated by the dashed white line. \\

\indent Figure~\ref{fig:CvsMu} shows the variation in steady state concurrence with temperature gradient (on the vertical axis) and applied voltage (on horizontal axis). The temperature of the left contact ($T_L$) is kept fixed at 10K while the right contact temperature ($T_R$) is varied between 10K to 20K, implying that the temperature gradient ($\Delta T = T_R - T_L$) varies between 0 to 10K. We observe that the two regions with non zero concurrence (= 1/3) correspond to $\ket{b1}$ or $\ket{b2}$ as the steady state. In between them, a mixture of $\ket{b1}$ and $\ket{b2}$ states is formed and net concurrence between the spin impurities comes out to be zero. It is evident that the entanglement between the two impurity spins can be effectively changed from 0 to 1/3 or vice versa by changing the applied voltage (keeping the contact temperatures constant), allowing the system to evolve in time undisturbed and waiting for it to reach the steady state. Post-selection can be done on those steady states whose concurrence is equal to 1/3 to obtain a maximally entangled state, as will be described later.
\subsection{Variation of Onsite Energy ($\epsilon$) and Exchange Coupling ($J$)}

\indent Figure~\ref{fig:Heatmap_w} illustrates the variation of the dot onsite energy along with change in the applied voltage. The temperature of both the contacts is held constant at 10K. There are two distinct sub-regions of non zero steady state concurrence corresponding to $\ket{b1}$ and $\ket{b2}$, separated by a short region around zero bias. The left region is triangular in shape as the concurrence is zero for all voltages when $\epsilon>2J$. This is because in such a case, the $\ket{b1}$ and $\ket{b2}$ states go above the zero electron occupancy states in energy (as evident from Fig.~\ref{fig:EnergySpectra}), and hence they do not arise as steady states for any applied voltage. The rectangular region for positive voltage bias is bounded on the right by $3J$, for voltages higher than this, the concurrence is always zero. \\

\indent Figure~\ref{fig:Heatmap_J} which represents variation of the exchange coupling and voltage again has two distinct sub-regions of non zero concurrence corresponding to $\ket{b1}$ and $\ket{b2}$. This time the positive voltage region with non zero concurrence is not rectangular as its right bound ($=3J$) increases as $J$ is increased.

\subsection{Maximal Entanglement Generation}
\indent Steady states of the system change with the tunable parameters as described previously. Partial entanglement exists between the ancillary qubits in the steady states $\ket{b1}$ and $\ket{b2}$. In this section we put forward a selection scheme to generate a maximally entangled state from a partially entangled steady state. We assume the system to be initially in some product state configuration given by $\rho_{D}\otimes \rho_{S_1}\otimes \rho_{S_2}$. Subsequently, when the contact voltage and other tunable parameters are fixed and the system is left undisturbed, it undergoes time evolution given by the positive map (as described by Eq.~\eqref{eq:L_master}). Under appropriate parametric conditions, this  map will generate a partially entangled steady state, such as $\ket{b1} = \sqrt{1/3}\ket{\uparrow} \otimes \ket{\Psi^{+}} - \sqrt{2/3}\ket{\downarrow} \otimes \ket{\uparrow \uparrow}$. This structure of $\ket{b1}$ suggests that a measurement can be performed on the quantum dot to extract the state of the ancilla qubits as well. Whenever the dot is found to be in a $\ket{\downarrow}$ state, the ancilla spins are left unentangled. However, if the dot is measured to be in an $\ket{\uparrow}$ state, the ancillae must be in $\ket{\Psi^{+}}$ state, which is a maximally entangled state. We can repeat the same conditions from the start until we have postselected for maximal entanglement. It can be inferred that the proposed device will generate maximal entanglement in the ancillae with a chance of one in three trials. To carry out the measurement based maximal entanglement generation scheme described above, it is necessary to readout the spin state inside the quantum dot. A quantum non-demolition measurement \cite{Tarucha} can be made by integrating dispersive read out into our setup. Using this, we can read out the spin state of the electron inside the quantum dot, from which the read out of the up-spin state generates a maximally entangled Bell pair among the impurities. Such a filtering procedure has been used in the context of entanglement thermal machines before \cite{tavakoli2018heralded}.

\section{Conclusion}
\indent Entanglement generation is an important problem for quantum technologies. In the context of quantum dot systems, such entanglement generation can be between quantum dots or in impurity atoms interacting with the quantum dots. The task of generating entanglement between quantum dots has been addressed by various techniques such as reservoir engineering, Floquet driving and strong coupling. Here we presented a solid state thermal machine based on quantum dots that is able to entangle distant spins in the steady state. We note that the steady state operation implies that the machine does not require active driving or other forms of time-dependent intervention. By analyzing the Liouvillian eigenspectrum as a function of the control parameters, we showed that our entanglement thermal machine operates over a large voltage region controlled by experimentally feasible steady state currents. Besides working with a non-zero voltage bias, the proposed device also works as a entanglement thermal machine under a temperature gradient that can even give rise to entanglement at zero voltage bias. Furthermore, simple post-selection schemes can be devised based on currently feasible non-demolition measurement techniques that can generate perfect Bell-pairs from the steady states using our generalized thermal machine. \\

\indent Quantum spins that are entangled via temperature or voltage gradients can serve as a platform for optical memories. Furthermore, such impurities embedded besides quantum dots in various architectures can serve as a conduit for quantum information. By employing experimentally straightforward operations such as SWAP gates and measurements, our technique of creating perfect Bell pairs can be used to grow a cluster state using well known techniques \cite{nielsen2006cluster}.\\

{\it{Acknowledgments:}} SV acknowledges support from a DST-SERB Early Career Research Award (ECR/2018/000957). BM and SV acknowledge the DST-QUEST grant number DST/ICPS/QuST/Theme-4/2019. Bitan De is supported by the National Science Centre (Poland) project
2019/35/B/ST2/00034. BM would like to acknowledge the Visvesvaraya Ph.D Scheme of the Ministry of Electronics and Information Technology (MEITY), Government of India, implemented by Digital India Corporation (formerly Media Lab Asia), and the Science and Engineering Research Board (SERB), Government of India, Grant No. STR/2019/000030. 
\bibliographystyle{apsrev4-1}
\color{RoyalBlue}

\end{document}